\documentclass[a4paper,10pt,twocolumn,superscriptaddress]{revtex4}
\usepackage{amsmath,amssymb}
\usepackage[final]{graphicx}
\usepackage{pdfpages}

\begin{document}

\title{Importance of interband transitions for the fractional quantum Hall effect in bilayer graphene}
\author{Kyrylo Snizhko}
\affiliation{Physics Department, Taras Shevchenko National University of Kyiv, Kyiv, 03022, Ukraine}
\affiliation{Department of Physics, Lancaster University, Lancaster, LA1~4YB, UK}
\author{Vadim Cheianov}
\affiliation{Department of Physics, Lancaster University, Lancaster, LA1~4YB, UK}
\author{Steven H. Simon}
\affiliation{Rudolf Peierls Centre for Theoretical Physics, Oxford University, Oxford, OX1~3NP, UK}

\begin{abstract}
Several recent works have proposed that electron-electron interactions in bilayer graphene can be tuned with the help of external parameters, making it possible to stabilize different fractional quantum Hall states. In these prior works, phase diagrams were calculated based on a single Landau level approximation. We go beyond this approximation and investigate the influence of polarization effects and virtual interband transitions on the stability of fractional quantum Hall states in bilayer graphene. We find that for realistic values of the dielectric constant, the phase diagram is strongly modified by these effects. We illustrate this by evaluating the region of stability of the Pfaffian state.
\end{abstract}

\maketitle

Recent experimental observation of the fractional steps in the Hall conductivity of ultraclean suspended graphene \cite{Du2009, Bolotin2009} and in graphene on the hexagonal boron nitride (h-BN) substrate \cite{Dean2011} has opened a new chapter in the physics of the fractional quantum Hall effect (FQHE) where new states of matter might be observed. For example, the peculiarities of the single-particle spectrum of graphene are predicted to result in novel strongly correlated Hall fractions such as the SU(4) generalization of the Halperin-Laughlin state \cite{Goerbig2007}. It has also been proposed that graphene allows for unprecedented tunability of the electron-electron interaction potential within a partially filled Landau level \cite{Abanin2011, Apalkov2011}, allowing access to the experimental investigation of quantum phase transitions between different topological orders \cite{Papic2011} and potentially stabilizing exotic topological states such as non-Abelian quantum Hall fluids \cite{Apalkov2011}. Of particular interest from this perspective is bilayer graphene (BLG) in which the matrix elements of the Coulomb interaction within one Landau level (also known as the Haldane pseudopotentials \cite{Haldane}) may be tuned {\it in situ} by the application of an external electric field breaking the symmetry between the two graphene layers \cite{Apalkov2011}.

The tunable quantum Hall effect in BLG was investigated theoretically in two recent papers \cite{Apalkov2011,Papic2011} where the stability conditions for several Abelian and non-Abelian FQHE states are found. Both groups use the same methodology based on the single Landau level approximation (SLLA). In this approximation the Hilbert space is constrained onto a given Landau level (LL) and the Hamiltonian takes the form of two-body interaction encoded in Haldane pseudopotentials \cite{Haldane}. The SLLA Hamiltonian is then exactly diagonalized for a small number of particles. The advantages of this approach are (a) its relative simplicity, and (b) its effectiveness in GaAs structures \cite{DasS} due to the large values of cyclotron gaps achieved in the experiment and a large value of the dielectric constant, which suppresses transitions between different LLs \cite{ChakrabortyBook}. However, as was discussed in \cite{Aleiner2011}, BLG is a narrow gap semiconductor in the strong coupling regime, where the effects of virtual interband transitions are essential. Moreover, as will be discussed in this work, the large cyclotron gap condition can be easily violated due to the peculiar properties of the single particle spectrum. In this paper we develop a theoretical framework incorporating such effects into the SLLA by means of perturbation theory. We are able to analyze the conditions of the applicability of our approach, and we find that over a broad range of parameters these interband transitions can completely change the effective interaction within this system. As an illustration of the importance of these effects, we use our approach to study the stability of the Moore-Read Pfaffian\cite{MooreRead} state in the BLG system. We find dramatic differences from the phase diagram predicted in Ref. \cite{Apalkov2011}. We conclude that all future analyses of the BLG system (or, indeed, all narrow gap semiconductors) will require more careful treatment of these effects than has been made in the past.

The single-particle Hamiltonian for BLG in a perpendicular electric field is \cite{MCF}
\begin{equation}
\label{Ham4}
H = \xi \left(
\begin{array}{cccc}
-U & 0 & 0 & v \pi^\dag\\
0 & U& v\pi& 0\\
0 & v\pi^\dag& U &\xi \gamma_1\\
v\pi & 0 &\xi\gamma_1 & -U
\end{array}
\right)
- 2 \mu_B B S_z
\end{equation}
where $\xi=\pm1$ for two different valleys, $\pi = p_x+ i p_y$, $p_i = -i \hbar \partial_i + e A_i/c $, $\vec{A}$ is a vector potential for the uniform perpendicular magnetic field $B,$ $2 U$ is the single-particle mini-gap tunable by the electric field (for simplicity we consider $U > 0$), $\mu_B$ is the Bohr magneton and $S_z$ is the z-projection of the electron spin. The Fermi velocity $v \approx 10^6\ \text{m/s}$, and the interlayer coupling is taken to be $\gamma_1 \approx 0.35\ \text{eV}$ \cite{Review}.

The eigenstates of the Hamiltonian \eqref{Ham4} with $|E| < \gamma_1$ are characterized by five quantum numbers: the valley index $\xi$, the LL number $n \in \mathbb{Z}_{+}$, the angular momentum $j_z = m + 1$ with $m \in (\mathbb{Z}_{+}-n)$, $s = \pm1$ which denotes whether the state has positive or negative energy and $S_z = \pm 1/2$. The wave functions in the $n$-th LL are:
\begin{equation}
\label{wavefunction}
\Psi_{n m}^{\xi s} = \left(
\begin{array}{c}
A_n^{\xi s} \psi_{n m}\\
B_{n}^{\xi s} \psi_{n-2, m+2}\\
C_{n}^{\xi s} \psi_{n-1, m+1}\\
D_{n}^{\xi s} \psi_{n-1, m+1}
\end{array}\right)
\end{equation}
where $\psi_{n m}$ is the nonrelativistic Landau wave function \cite{Landau} in the $n$-th LL (negative LL wave functions are identically zero by definition). The amplitudes $A_{n}^{\xi s}, B_{n}^{\xi s}, C_{n}^{\xi s}, D_{n}^{\xi s}$ depend on $U$ and $B$, and this dependence is crucial for tuning of the interaction matrix elements. The single particle spectrum can be found from the equation
\begin{equation}
\label{spectreq}((u-\xi \varepsilon)^2-\gamma^2 n)((u+\xi \varepsilon)^2-\gamma^2 (n-1)) = \gamma^4 (\varepsilon^2-u^2)
\end{equation}
where, following \cite{MCF}, we introduce $\omega_c = 2 v^2 e B/\gamma_1 c$; we also introduce dimensionless parameters $u = U/(\hbar \omega_c)$, $\gamma^2 = \gamma_1/(\hbar \omega_c)$, $\varepsilon = (E + 2 \mu_B B S_z)/(\hbar \omega_c)$, where $E$ is the energy.

\begin{figure}[tbp]
\centering
\includegraphics[width=.96\columnwidth]{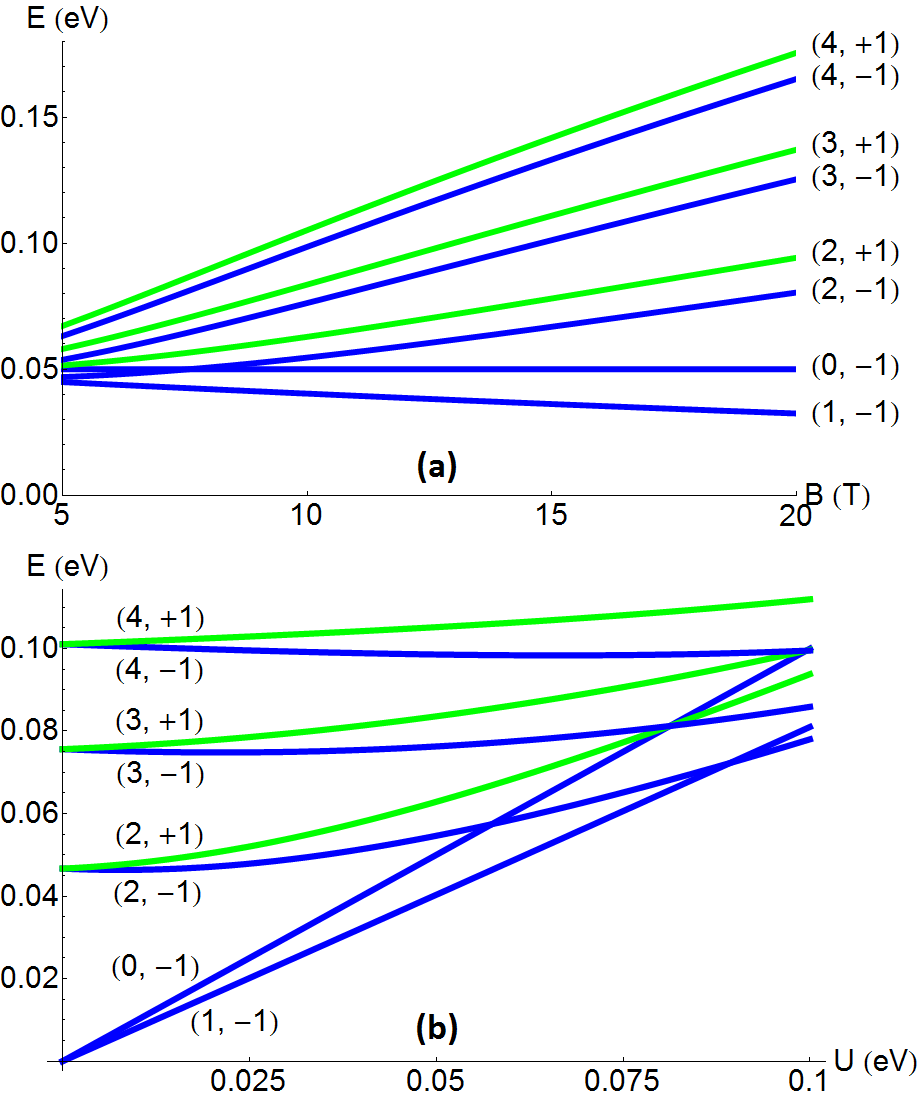}
\caption{(color online). The dependence of the lowest LLs' energies on (a) the magnetic field $B$ at $U = 50\ \text{meV}$, and (b) on the gap parameter $U$ at $B = 10\ \text{T}$. Each level is labeled by a pair of quantum numbers $(n, \xi)$.}
\label{fig:spectr}
\end{figure}

The most quantitative method for the theoretical investigation of the FQHE is exact diagonalization of the Hamiltonian of a small particle number system. However, this is possible only for relatively small Hilbert spaces, so going beyond the SLLA makes calculations practically impossible. The peculiarities of the BLG single-particle spectrum put rather tough constraints on the applicability of the SLLA. Figure \ref{fig:spectr}a shows the dependence of the several lowest LLs' energies on the magnetic field at $U = 50\ \text{meV}$. Only the positive energy part of the spectrum is shown. The negative energy LLs can be obtained by the particle-hole conjugation. We label each positive-energy LL by a pair of quantum numbers $(n, \xi)$. One can see that at high magnetic fields the levels group into quasidegenerate doublets separated by energy of the order of $\hbar \omega_c$. Figure \ref{fig:spectr}b shows the dependence of the same LL energies on the gap parameter $U$ at $B = 10\ \text{T}$. Note that at large enough $U$ (or small enough $B$) LLs cross. Thus the applicability of the SLLA puts an upper limit on $U$ at any given $B$. At the same time, due to the small energy separation between the levels in each doublet (which is proportional to $U$) the SLLA is valid only for large enough $U$.

Next we discuss the effect of the single-particle band structure on the interaction physics of FQHE in BLG. One can safely work in the approximation neglecting non-conservation of the spin and valley quantum numbers. In this approximation the following effects can be important. First, BLG is a narrow-gap semiconductor with strong vacuum polarization effects \cite{Aleiner2011}. Second, the interaction can lead to the change of the order of filling of the LLs or to the appearance of the spin and valley unpolarized states. Third, even when the dielectric constant is large virtual hopping between LLs can still renormalize the intra-LL interaction.

\begin{figure}[tbp]
\centering
\includegraphics[width=.96\columnwidth]{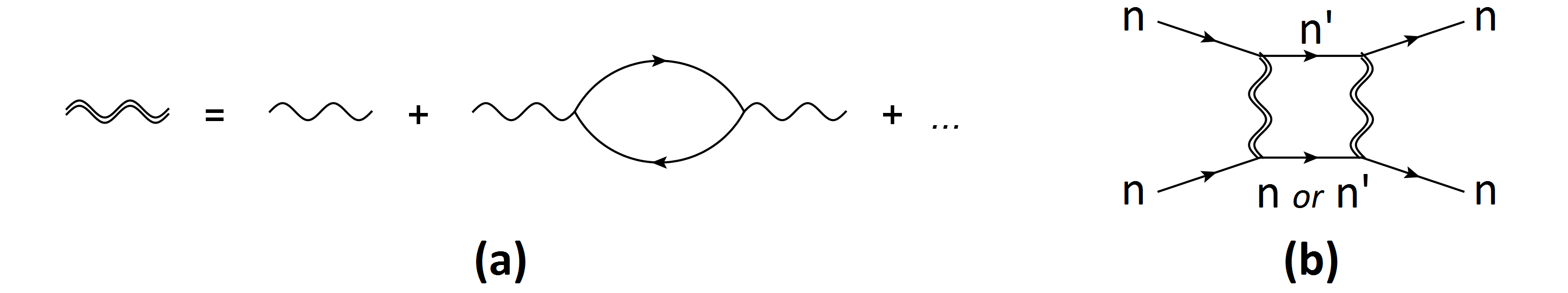}
\caption{Feynman diagrams\cite{JaxoDraw} showing renormalization of the electron-electron interaction due to (a) the vacuum polarization processes, and (b) the simplest processes involving virtual hopping of one or both of the two interacting electrons from the $n$-th LL to the $n'$-th LL.}
\label{fig:diag}
\end{figure}

\textit{Vacuum polarization}. The virtual processes shown in fig.~\ref{fig:diag}a lead to a renormalization of the electron-electron interaction potential. The Fourier transform of the renormalized interaction potential can be represented as follows\cite{Morf}:
\begin{equation}
\label{screening}V_{scr}(q) = \frac{v(q)}{1 + v(q) \Pi(q, \omega = 0)}
\end{equation}
where $v(q) = 2\pi e^2/(q \kappa)$ is the Fourier transform of the bare Coulomb potential, $\kappa$ is the dielectric constant of the environment\cite{foot1} and $\Pi(q, \omega)$ is the polarization function. The approximation neglecting the retardation effects ($\omega = 0$) is justified as long as we are interested in energies much smaller than the inter-LL gaps. We take the polarization function from the random-phase approximation calculation for BLG with magnetic field, which can be justified within the $1/N$ expansion \cite{Aleiner2011} ($N = 2 \text{ spin projections} \times 2 \text{ valleys} = 4$). Since $\Pi(q, \omega = 0) \propto q^2$ screening is not efficient at large distances, however it strongly affects the first few Haldane pseudopotentials (corresponding to distances of the order of magnetic length) which have the most significant impact on the stability of any FQHE state.

\textit{Population reversal of LLs}. The order of levels in Fig. \ref{fig:spectr} prescribes the natural order of filling of the LLs by electrons in the independent electrons approximation. However, we find that for integer filling fractions the electron-electron interaction leads to a reversal of this natural order in a significant part of parameter space. For example, the Coulomb energy of the fully filled $(2, +1)$ LL is less than the one of the fully filled $(2, -1)$ for $U > 0$. In the region where the interaction is strong compared to the gap between the two levels this leads the fully filled $(2, +1)$ LL having lower total energy than the fully filled $(2, -1)$ level. Thus the former will be filled before the latter. When the quasidegenerate levels are from different valleys, valley-unpolarized state can be preferred, particularly for fractional filling. Furthermore, when the quasidegenerate levels are from the same valley (as in the $n = 0$ and $n = 1$ case) level mixing can occur. These are interesting effects which are, however, beyond the scope of this paper. Whereas for fractional filling such effects are much more difficult to analyze, population reversal at the integer filling fraction is an indicator of strong violation of the SLLA. Thus we constrain our analysis to the region of the parameter space where no population reversal occurs at integer filling. This guarantees that the states are valley polarized. For the valley polarized states one can investigate whether the state is spin-polarized. Generally, spin-unpolarized states are not favored by Coulomb repulsion unless the potential is hollow-core, which we find to be not true in our case.

\textit{Renormalization of pseudopotentials due to virtual hopping}. The SLLA is exact in the limit of infinite energy difference $\Delta E$ between LLs. For finite $\Delta E$, the SLLA pseudopotentials acquire corrections due to virtual transitions between the LLs such as, for example, shown in Fig. \ref{fig:diag}b\cite{foot2}. Such corrections are theoretically tractable only in the perturbative regime (when they are small), however even the presence of small corrections may dramatically affect the phase diagram due to the extreme sensitivity of the FQHE states to the details of the interaction. In this work we take the virtual hopping corrections into account in second order perturbation theory (Fig. \ref{fig:diag}b). We restrict the region of validity of our consideration by requiring the third order corrections to be smaller than the second order ones.

To investigate the role of these effects on the stability of FQHE states we focus on the Pfaffian state. Our choice is motivated by the following considerations. First, this state is particularly sensitive to the details of the interaction so it is a good illustration for our analysis. Second, the stability of this state in BLG was investigated in Refs.~\cite{Apalkov2011, Abanin2011} in the SLLA approximation but without these effects taken into account, so we can compare the phase diagrams. Third, the Pfaffian itself is an important state because it is an example of the non-Abelian topological fluid. The tunable parameters are the magnetic field $B$, the electric field which determines the mini-gap parameter $U$ and the dielectric constant $\kappa$ which controls the deviation from the naive SLLA (which is exact for $\kappa \rightarrow \infty$). We can also choose the half-filled LL number. Here we will concentrate only on the two levels: $(1, -1)$ and $(2, -1)$. The $(1, -1)$ level wave function is constructed from the nonrelativistic $n = 0$ and $n = 1$ LL wave functions, the $(2, -1)$ level wave function is constructed from the nonrelativistic $n = 0,1,2$ LL wave functions. In both cases one can tune the pseudopotentials close to their values at the nonrelativistic $n = 1$ LL, where the $5/2$ state is observed in GaAs.

The tuning mechanisms are, however, different for the two levels. Amplitudes of the wave function \eqref{wavefunction} in the $(1, -1)$ LL show little dependence on $U$ so the main control parameter is $B$. In contrast, the amplitudes of the wave function in the $(2, -1)$ LL mainly depend on one parameter which is the $U/\hbar \omega_c$ ratio, so both $B$ and $U$ can be used for tuning.

The main factors determining deviation from the naive SLLA for the two levels are the polarization and virtual hopping to the nearby levels. For the $(1, -1)$ LL this is hopping to the $(0, -1)$ and the $(2, -1)$ LLs, while for the $(2, -1)$ LL the important hopping is to the $(3, -1)$ LL. In addition to this, for the $(2, -1)$ LL, it is important to consider effects of mixing with the $(0, -1)$ and possible population order reversal with $(2, +1)$. The latter are important factors restricting the region of applicability of perturbative analysis, however, when suppressed they do not lead to a renormalization of the intra LL interaction.

\begin{figure}[tbp]
\centering
\includegraphics[width=.96\columnwidth]{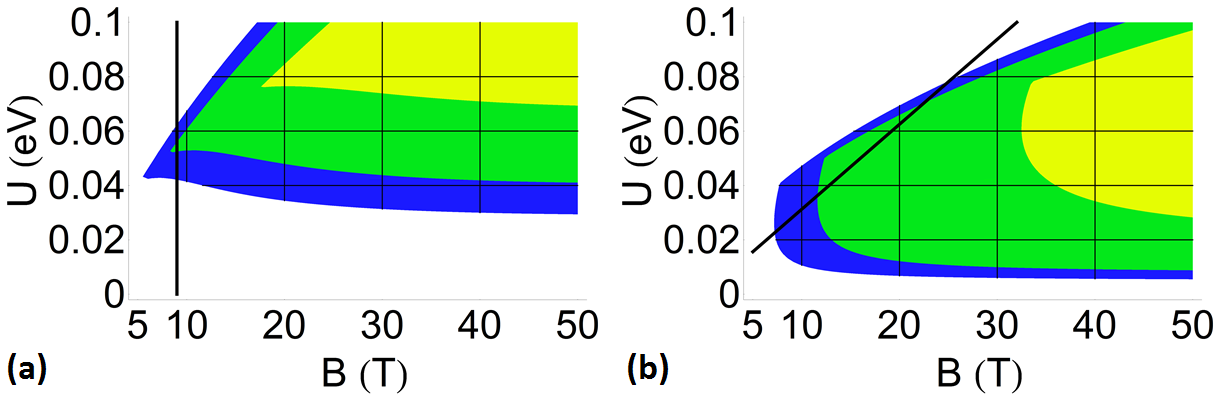}
\caption{(color online). The region ($(a)$ -- for the $(1, -1)$ LL, $(b)$ -- for the $(2, -1)$ LL) of the applicability of perturbative analysis for fixed values of $\kappa = 5, 10\text{ and }15$. The size of the region increases with increasing $\kappa$. The thick black line shows where the maximum overlap with the Moore-Read Pfaffian for the bare Coulomb interaction is achieved.}
\label{fig:zones}
\end{figure}

Figures \ref{fig:zones}a and \ref{fig:zones}b show the regions of the applicability of perturbative analysis for different values of $\kappa$ for the $(1, -1)$ and the $(2, -1)$ LLs respectively. For the $(1, -1)$ LL the region is bounded from above by the condition of small hopping to the $(2, -1)$ LL, the lower bound is due to the condition of small hopping to the $(0, -1)$ LL. At small enough magnetic fields, at least one of the conditions is violated at all values of $U$. For the $(2, -1)$ LL the region's upper bound is due to the condition of small mixing with the $(0, -1)$ LL, while the lower and the left bounds are due to the condition of absence of the population reversal. The thick black line shows where the maximum overlap with the Pfaffian for the bare Coulomb interaction is achieved. One can see that for small dielectric constants this line lies outside the region of validity of perturbative analysis; however, for large enough $\kappa$ they intersect near $U = 50\ \text{meV}$ in both cases. This happens at $\kappa \gtrsim 10$ and $\kappa \gtrsim 6$ for the $(1, -1)$ LL and the $(2, -1)$ LL respectively.

\begin{figure}[tbp]
\centering
\includegraphics[width=.96\columnwidth]{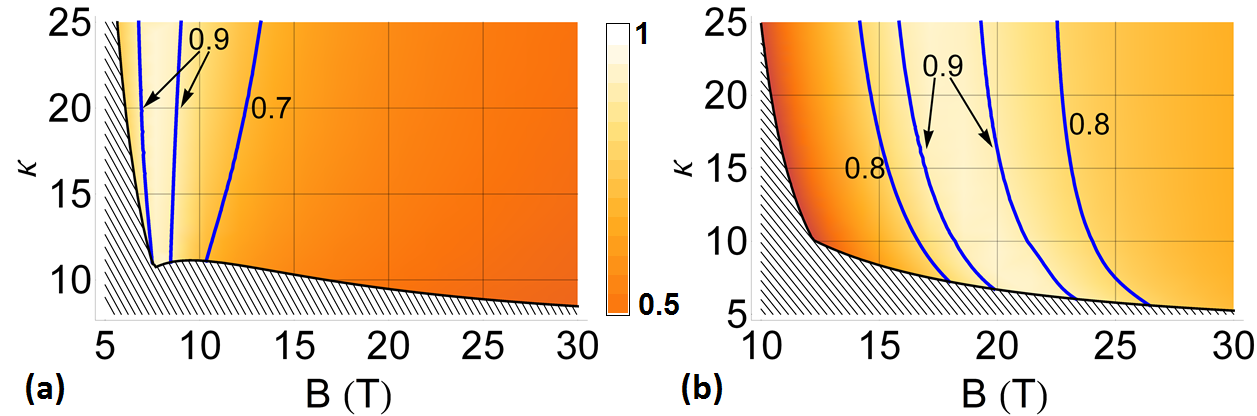}
\caption{(color online). Color plot of the overlap of the ground state with the Moore-Read Pfaffian for $12$ particles at $U = 50\ \text{meV}$ as a function of the magnetic field $B$ and the dielectric constant $\kappa$ ($(a)$ -- for the $(1, -1)$ LL, $(b)$ -- for the $(2, -1)$ LL). Contours show the lines of constant overlap. The region where perturbative analysis is not applicable is hatched.}
\label{fig:overlap}
\end{figure}

Figures \ref{fig:overlap}a and \ref{fig:overlap}b show the dependence of the overlap of the exact ground state of the system with the Pfaffian on the magnetic field and the dielectric constant at $U = 50\ \text{meV}$ for the $(1, -1)$ and the $(2, -1)$ LLs respectively. The region where perturbative analysis is not applicable is hatched. As one can see, for the $(1, -1)$ level, a high overlap up to $0.94$ (compare with non-relativistic $n = 1$ level overlap of $0.7$) is achieved for all admissible values of $\kappa$ near $B = 8\ \text{T}$. For the $(2, -1)$ level a high overlap up to $0.93$ is achieved near $B = 20\ \text{T}$, also for all admissible values of $\kappa$. However, the behavior of the high-overlap region is different for the two cases. The region is situated at the same magnetic field but becomes narrower with decreasing $\kappa$ for the $(1, -1)$ LL. For the $(2, -1)$ LL the region also gets narrower with decreasing $\kappa$ but its position shifts to higher values of the magnetic field.

The authors of \cite{Apalkov2011} found that in the $(1, -1)$ LL, high overlap is achieved in the region near $B = 10\ \text{T}$. We find that the region of high overlap is situated there for large enough values of $\kappa$. However, for smaller values of $\kappa \lesssim 10$ the effect of level mixing becomes significant which makes observation of the Pfaffian state unlikely. The $(2, -1)$ LL was also considered in \cite{Apalkov2011} where it was concluded that the maximal overlap with the Pfaffian on this level is less than $0.6$. Our results do not support this conclusion (even for the bare Coulomb interaction).

The previous consideration shows that BLG can be tuned into the regime of high overlap with the Pfaffian; however, one needs a higher dielectric constant than the usual $\kappa \approx 2.5$ for graphene on a $\rm{SiO_2}$ substrate. This is experimentally achievable. For example, on a $\rm{HfO_2}$ substrate \cite{Hafnium} $\kappa$ is around $12.5$. For this value of $\kappa = 12.5$ the required magnetic fields (around $B = 10\ \text{T}$ for the $(1, -1)$ LL and around $B = 20\ \text{T}$ for the $(2, -1)$ LL) are quite realistic. The gap to the first excited state at these parameter values is around $2\ \text{K}$ and $8\ \text{K}$ for the $(1, -1)$ and the $(2, -1)$ LLs, respectively. With increasing magnetic field we find that the gap monotonically increases to the values of around $20\ \text{K}$ and $17\ \text{K}$, respectively, at $B = 30\ \text{T}$. At the same time the overlap decreases to around $0.5$, which is still fairly large.\cite{foot3} This result, obtained for a finite number of particles, suggests that the system may still be in the same topological phase at higher magnetic fields.

To conclude, we have analyzed the influence of inter-Landau level transitions on the phase diagram of the FQHE states. We find that the SLLA can only be used under quite stringent restrictions. For moderate values of the dielectric constant these effects can be taken into account perturbatively leading to some modification of the phase diagram, with dramatic modification outside the perturbative approach applicability region where Pfaffian observation is unlikely.

We thank Prof. V. Fal'ko for useful discussions. We thank the authors of DiagHam code for numerical diagonalization, especially N. Regnault. We also thank the authors of the JaxoDraw program, which was used to draw the Feynman diagrams.

\clearpage



\section*{Supplementary material}

# Importance of interband transitions for the fractional quantum Hall effect in bilayer graphene — Supplementary Material

Kyrylo Snizhko,[1,2] Vadim Cheianov,[2] and Steven H. Simon[3]

[1]*Physics Department, Taras Shevchenko National University of Kyiv, Kyiv, 03022, Ukraine*
[2]*Department of Physics, Lancaster University, Lancaster, LA1 4YB, UK*
[3]*Rudolf Peierls Centre for Theoretical Physics, Oxford University, Oxford, OX1 3NP, UK*

In this section we provide additional details regarding the stability of the Pfaffian state.

Figures 1a and 1b show the gap between the exact ground state of the system with 12 particles and its exact first excited state. In these figures, the gap is plotted as a function of the magnetic field and the dielectric constant at $U = 50$ meV for the $(1, -1)$ LL (Fig. 1a) and the $(2, -1)$ LL (Fig. 1b). The region where perturbative analysis is not applicable is hatched. Figures 2a and 2b show the same gap in the units of typical Coulomb energy $e^2/\kappa\ell$ where $\ell$ is the magnetic length $\ell = \sqrt{\hbar c/eB}$.

As one can see, for both levels gaps increase as the magnetic field increases (staring from points where maximum overlap occurs) and as the dielectric constant decreases. This can be partially (for magnetic field) or fully (for the dielectric constant) attributed to the increase of $e^2/\kappa\ell$. This result, obtained for a finite number of particles, suggests that the system may still be in the same topological phase at higher magnetic fields. The same holds for low dielectric constants down to boundary of the perturbative approach applicability region: for smaller dielectric constants the physics is essentially non-single Landau level.

Figures 3a and 3b show the dependence of the gap on the magnetic field at $U = 50$ meV and $\kappa = 12.5$ for the $(1, -1)$ and the $(2, -1)$ LLs respectively. Figures 4a and 4b show the same data plotted in the units of $e^2/\kappa\ell$. One peculiarity may be worth noticing: the gap has a local minimum near the maximum overlap point.

Figures 5a and 5b show the dependence of the gap on the dielectric constant at $U = 50$ meV and fixed $B$ for the $(1, -1)$ and the $(2, -1)$ LLs respectively ($B = 8$ T for $(1, -1)$ LL and $B = 19$ T for $(2, -1)$ LL). Figures 6a and 6b show the same data plotted in the units of $e^2/\kappa\ell$. This illustrates once more that the gaps increase

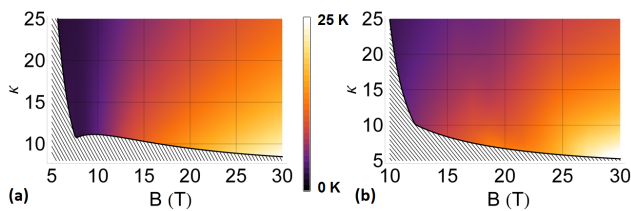

FIG. 1: Color plot of the gap between the ground state and the first excited state computed for 12 particles at $U = 50$ meV as a function of the magnetic field $B$ and the dielectric constant $\kappa$ ((a) – for the $(1, -1)$ LL, (b) – for the $(2, -1)$ LL). The region where perturbative analysis is not applicable is hatched.

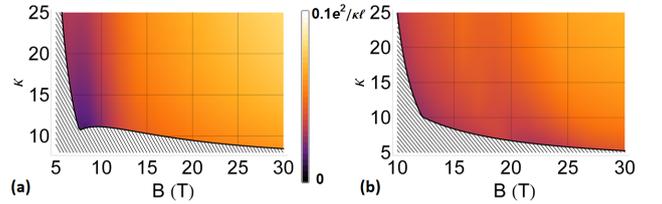

FIG. 2: Color plot of the gap between the ground state and the first excited state computed for 12 particles at $U = 50$ meV as a function of the magnetic field $B$ and the dielectric constant $\kappa$ ((a) – for the $(1, -1)$ LL, (b) – for the $(2, -1)$ LL). The region where perturbative analysis is not applicable is hatched.

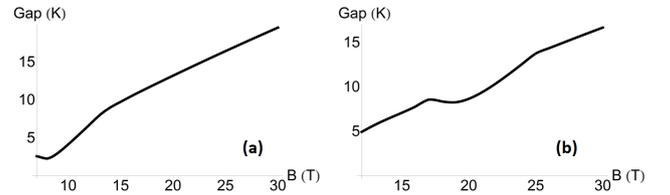

FIG. 3: Dependence of the gap between the ground state and the first excited state computed for 12 particles at $U = 50$ meV and $\kappa = 12.5$ as a function of the magnetic field $B$ ((a) – for the $(1, -1)$ LL, (b) – for the $(2, -1)$ LL). Only the part where perturbative analysis is applicable is shown.

as the magnetic field increases and as the dielectric constant decreases, however, the "intrinsic" gaps (gaps in units of $e^2/\kappa\ell$) increase as the magnetic field increases but decrease as the dielectric constant decreases. This shows the tendency towards the loss of stability by the Pfaffian state at small dielectric constants.

Figure 7 compares the data on the overlap with Pfaf-

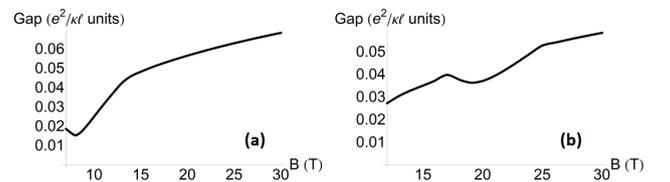

FIG. 4: Dependence of the gap between the ground state and the first excited state computed for 12 particles at $U = 50$ meV and $\kappa = 12.5$ as a function of the magnetic field $B$ ((a) – for the $(1, -1)$ LL, (b) – for the $(2, -1)$ LL). Only the part where perturbative analysis is applicable is shown.

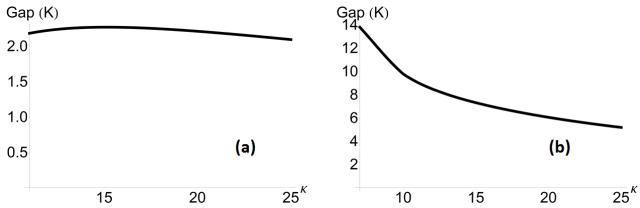

FIG. 5: Dependence of the gap between the ground state and the first excited state computed for 12 particles at $U = 50$ meV and fixed $B$ as a function of the dielectric constant $\kappa$ ((a) – for the $(1,-1)$ LL at $B = 8$ T, (b) – for the $(2,-1)$ LL at $B = 19$ T). Only the part where perturbative analysis is applicable is shown.

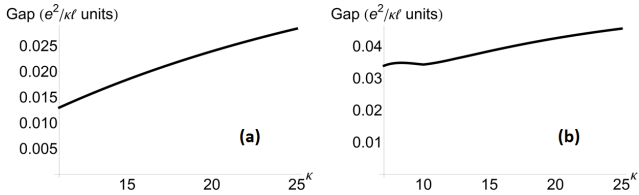

FIG. 6: Dependence of the gap between the ground state and the first excited state computed for 12 particles at $U = 50$ meV and fixed $B$ as a function of the dielectric constant $\kappa$ ((a) – for the $(1,-1)$ LL at $B = 8$ T, (b) – for the $(2,-1)$ LL at $B = 19$ T). Only the part where perturbative analysis is applicable is shown.

fian and the gap to the first excited state for the $(1,-1)$ and the $(2,-1)$ LLs in BLG and the non-relativistic $n = 1$ LL. The data are shown for different numbers of particles $N = 8, 10, 12, 14$. For the $(1,-1)$ and the $(2,-1)$ LLs in BLG external parameters are set to be near the maximum overlap region. The gap is presented in units of the typical Coulomb energy $e^2/\kappa\ell$. The results for BLG levels appear to be more stable as the number of particles grows than for the non-relativistic system.

Figures 8, 9 present the dependence of the overlap and the gap at several parameter points for $(1,-1)$ LL and $(2,-1)$ LL respectively. Points are taken to be at the same mini-gap and the dielectric constant values as in

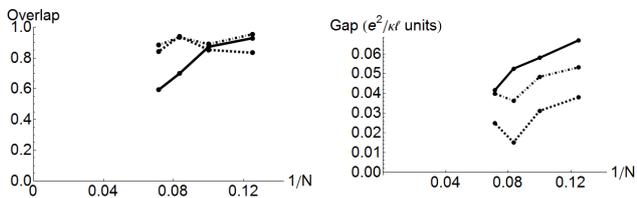

FIG. 7: Dependence of the overlap with Pfaffian and gap to the first excited state on the number of particles $N$ ($N = 8, 10, 12, 14$). Solid line is for non-relativistic $n = 1$ LL. Dashed line is for $(1,-1)$ LL at $U = 50$ meV, $\kappa = 12.5$ and $B = 8$ T. Dot-dashed line is for $(2,-1)$ LL at $U = 50$ meV, $\kappa = 12.5$ and $B = 19$ T. The gap is presented in units of typical Coulomb energy $e^2/\kappa\ell$.

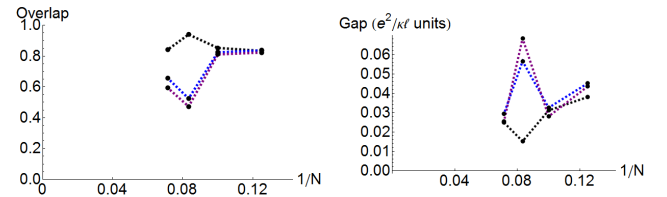

FIG. 8: Dependence of the overlap with Pfaffian and gap to the first excited state on the number of particles $N$ ($N = 8, 10, 12, 14$) for $(1,-1)$ LL at $U = 50$ meV, $\kappa = 12.5$ and $B = 8$ T (black), $B = 20$ T (blue), $B = 30$ T (purple). The gap is presented in units of typical Coulomb energy $e^2/\kappa\ell$.

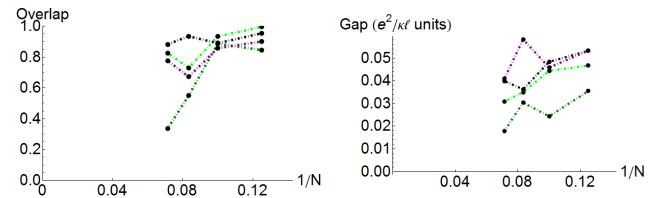

FIG. 9: Dependence of the overlap with Pfaffian and gap to the first excited state on the number of particles $N$ ($N = 8, 10, 12, 14$) for $(2,-1)$ LL at $U = 50$ meV, $\kappa = 12.5$ and $B = 19$ T (black), $B = 13$ T (dark green), $B = 15$ T (green), $B = 30$ T (purple). The gap is presented in units of typical Coulomb energy $e^2/\kappa\ell$.

the previous figure, but the magnetic field changes. For $(1,-1)$ LL we took the magnetic fields to be only higher than the maximum overlap region (since lower fields are outside the region of applicability of perturbative approach). For $(2,-1)$ LL we took both higher and lower magnetic fields. As one can see the overlap at the maximum point seems to be more stable as system size increases than at the other points. However, the overlap at higher magnetic field for $(2,-1)$ LL is still quite high and quite stable. The situation with the gaps appears similar. Though this does not give any conclusive answer, one might suggest that for $(1,-1)$ LL the Pfaffian state is stable in the neighbourhood of the maximum overlap region, while for $(2,-1)$ LL it may be stable also at higher magnetic fields.




\begin{thebibliography}{100}
\bibitem{foot1} The dielectric constant which BLG feels is $\kappa = (\kappa_1+\kappa_2)/2$, where $\kappa_1$ and $\kappa_2$ are the dielectric constants of the environment below and above the sheet of BLG.
\bibitem{foot2} In fact, virtual transitions also lead to the appearance of effective many-body interactions, however these are strongly suppressed in the perturbative regime.
\end{thebibliography}
\end{document}